\documentstyle[preprint,aps,epsfig,amsbsy,floats]{revtex}
 \tightenlines

\begin{document}
\newcommand{\beq}{\begin{equation}}
\newcommand{\eeq}{\end{equation}}
\newcommand{\beqa}{\begin{eqnarray}}
\newcommand{\eeqa}{\end{eqnarray}}
\newcommand{\sr}{\sqrt}
\newcommand{\fr}{\frac}
\newcommand{\mn}{\mu \nu}
\newcommand{\G}{\Gamma}

\draft \preprint{astro-ph/0408278,~ INJE-TP-04-06}
\title{Cosmological constraints on inflation potential:
the role of Gauss-Bonnet  corrections in
 braneworld scenarios}
\author{Kyong Hee Kim and  Yun Soo Myung\footnote{E-mail address:
ysmyung@physics.inje.ac.kr}}
\address{
Relativity Research Center and School of Computer Aided Science\\
Inje University, Gimhae 621-749, Korea} \maketitle

\begin{abstract}
We calculate the spectral index and tensor-to-scalar ratio for
patch inflation defined by $H^2\approx \beta^2_q V^q$ and
$\dot{\phi}\approx -V'/3H$, using the slow-roll expansion. The
patch cosmology arisen from the Gauss-Bonnet braneworld  consists
of Gauss-Bonnet (GB), Randall-Sundrum (RS), and 4D general
relativistic (GR) cosmological models. In this work, we choose
large-field potentials of $V=V_0\phi^p$ to compare with the
observational data. Since second-order corrections are rather
small in the slow-roll limit, the leading-order calculation is
sufficient to compare with the data. Finally, we show that it is
easier to discriminate between quadratic potential and quartic
potential in the GB cosmological model rather than the GR or RS
cosmological models.
\end{abstract}

\thispagestyle{empty}
\setcounter{page}{0}
\newpage
\setcounter{page}{1}
\section{introduction}
There has been much interest in the phenomenon of localization of
gravity proposed by Randall and Sundrum (RS)~\cite{RS2}. They
assumed a 3-brane with a  positive-tension in the five dimensional
(5D) anti-de Sitter spacetime. They obtained a localized gravity
on the brane by fine-tuning the tension  to the cosmological
constant. Recently, several authors studied the cosmological
implications of a brane world scenario. The brane cosmology
contains some important deviations from the
Friedmann-Robertson-Walker (FRW) cosmology\cite{BDL1,BDL2}.
Especially, the Friedmann equation is modified in the high-energy
regime significantly.

On the other hand, it is generally accepted that curvature
perturbations produced during inflation are considered to be
 the origin of  inhomogeneities necessary for
explaining  cosmic microwave background (CMB) anisotropies and
large-scale structures. The latest results come from  Wilkinson
Microwave Anisotropy Probe (WMAP)\cite{Wmap1}, Sloan Digital Sky
Survey (SDSS)\cite{SDSS} and others.  These put forward more
constraints on cosmological models and confirm the emerging
standard model of cosmology, a flat $\Lambda$-dominated universe
seeded by scale-invariant adiabatic Gaussian fluctuations. In
other words, these results coincide with theoretical  predictions
of the slow-roll inflation based on general relativity with a
single inflaton.

If the brane inflation occurs, one expects that it provides us
different results in the high-energy
regime\cite{MWBH,LT,Char,RL,TL}. In the slow-roll
approximation\cite{SL,STG}, there is no significant change in the
power spectrum  between the standard and brane cosmology up to
first-order corrections\cite{Cal1}.  In order to distinguish
between the standard and brane inflation apparently, it is
necessary to calculate their power spectra up to second-order
using the slow-roll expansion\cite{SG}.  Unfortunately, it is not
easy to discriminate between the standard  and brane inflation
because second-order corrections are rather small in the slow-roll
limit\cite{KLM1}. Furthermore,  the degeneracy  exists between
scalar and tensor perturbation. This is expressed as the
consistency relation of $R=-8n_T$ in the standard inflation. This
 relation remains unchanged even in the brane cosmology\cite{HL,Cal2}.
In order to handle this degeneracy problem, the authors
in\cite{DLMS} calculated the tensor spectrum in the Gauss-Bonnet
braneworld. They found that this  relation could be broken by the
Gauss-Bonnet term. However, this breaking  is so mild that the
likelihood values for the Gauss-Bonnet brane inflation are nearly
identical  to those in the standard inflation\cite{TSM}. Thus it
seems that the introduction of a Gauss-Bonnet term in the
braneworld could not distinguish between the standard inflation
and brane inflation.

In this work, we study whether or not the quartic potential
($V=V_0\phi^4$) could be  ruled out  in view of the patch
inflation defined by $H^2\approx \beta^2_q V^q, \dot{\phi}\approx
-V'/3H$. We note that the patch cosmology  comes from the
Gauss-Bonnet braneworld. First, in order to see the difference
between different models, we calculate the scalar spectral index
up to second-order using the slow-roll expansion. Since
second-order corrections are rather small in the slow-roll limit,
their theoretical points are not significantly moved from the
leading-order results.  Thus we use the leading-order spectral
index and tensor-to-scalar ratio to select  which model is mostly
suitable for discriminating between quadratic potential and
quartic potential. As a result,  the GB model is turned out to be
the best one  to choose
 a favorable potential of $V=V_0\phi^2$ and to reject an
 unfavorable potential of $V=V_0\phi^4$.

The organization of this paper is as follows. In Section II we
briefly review the patch cosmology and  relevant cosmological
parameters.   We choose large-field potentials to compare the
theoretical points with the observation data in Section III.
Finally we discuss our results in Section IV. We list the
second-order corrections to cosmological parameters in Appendix A
and their explicit forms for large-field potentials in Appendix B.

\section{patch cosmology}
We start with an effective  Friedmann equation arisen from the
Gauss-Bonnet braneworld  by adopting a flat FRW metric as the
background spacetime on the brane\footnote{For reference,  we add
the action for the  Gauss-Bonnet braneworld scenario\cite{Char}:
$S=\fr{1}{2\kappa^2_5}\int_{{\rm
bulk}}d^5x\sqrt{-g_5}\Big[R-2\Lambda_5\Big(R^2-4R_{\mu\nu}R^{\mu\nu}+
R_{\mu\nu\rho\sigma}R^{\mu\nu\rho\sigma}\Big)\Big]+\int_{{\rm
brane}}d^4x \sqrt{-g}\Big[-\lambda +{\cal L}_{{\rm matter}} \Big]$
with $\Lambda_5=-3\mu^2(2-\beta)$ for an AdS bulk cosmological
constant and ${\cal L}_{{\rm matter}}$ for inflation on the brane.
Its full Friedmann equation is given by a complicated form
$2\mu(1+H^2/\mu^2)^{1/2}\Big[3-\beta+3\beta
H^2/\mu^2\Big]=\kappa_5^2(\rho+\lambda)$.}\cite{Cal1,DLMS,TSM}
\begin{equation}
\label{Heq} H^2=\beta^{2}_{q}\rho^{q},
\end{equation}
where $H=\dot{a}/a$ is the Hubble parameter,  $q$ is a parameter
describing each  cosmological model, and $\beta_{q}^{2}$ is a
factor with energy dimension $[\beta_q]=E^{1-2q}$.  An additional
parameter defined by $\theta=2(1-1/q)$ is introduced for
convenience. We call the above defined on the $q$-dependent energy
regimes as a whole ``patch cosmology". We summarize relevant
cosmological models and their parameters in the patch cosmology in
Table I. $\kappa^{2}_{5}=8\pi M_{5}^{-3}(\kappa^2_4=8\pi
M_4^{-2})$ is the 5D (4D) gravitational coupling constant.
$\alpha$ is the Gauss-Bonnet coupling which may be related to the
string energy scale $g_s$ as $\alpha=1/8g_s$ and  $\lambda$ is a
brane tension. The two relations between these are
$\kappa^2_4/\kappa^2_5=\mu/(1+\beta)$ and $
\lambda=2\mu(3-\beta)/\kappa_5^2$, where $\beta=4\alpha\mu^2\ll
1,~\mu=1/\ell$ with  an AdS$_5$ curvature radius $\ell$.  The
Randall-Sundrum case of $\mu=\kappa_4^2/\kappa_5^2$ is recovered
when $\alpha=0$.
\begin{table}
 \caption{Three relevant models for the patch cosmology and their parameters classifying different models.}
 \begin{tabular}{lp{4.5cm}p{3cm}}
 model   & $q~(\theta)$ & $\beta^2_q$ \\ \hline
 GB      & 2/3 ($-1$)   & $(\kappa^2_5/16\alpha)^{2/3}$\\
 RS      & 2 (1)       & $\kappa^2_4/6\lambda$ \\
 GR      & 1 (0)          & $\kappa_4^2/3$
 \end{tabular}
 \end{table}
Introducing an inflaton $\phi$ confined to the brane, one finds
the equation
\begin{equation}
\label{seq} \ddot{\phi}+3H\dot{\phi}=-V^{\prime},
\end{equation}
where the dot and prime denote  the derivative with respect to
time $t$ and $\phi$, respectively. Its energy density and pressure
are given by $\rho=\dot{\phi}/2+V$ and $p=\dot{\phi}/2-V$. From
now on we use the slow-roll formalism for inflation: an
accelerated universe $(\ddot a>0)$ is being driven by a single
scalar field slowly rolling down its potential toward a local
minimum. This means that Eqs.(\ref{Heq}) and (\ref{seq}) take the
approximate forms:
\begin{equation}
H^2\approx \beta^2_q V^q,~ \dot{\phi}\approx -V'/3H.
\end{equation}
This implies that the cosmological acceleration is  given by a
fluid with a vacuum-like equation of state $p \approx -\rho$. If
$p=-\rho$ with $\dot{\phi}=0$, it corresponds to a de Sitter
inflation with $a(t)=a_0e^{Ht}$. We note that this is  the only
background to obtain gravitational waves from the braneworld
scenario\cite{LMW}. In order for an inflation to end and the
universe to transfer to the radiation-dominated universe, we need
the  slow-roll mechanism. For this purpose, we introduce Hubble
slow-roll parameters ($\epsilon_1,~\delta_n$) on the brane as
\begin{equation}
\epsilon_1 \equiv -\frac{\dot H}{H^2}\approx \epsilon_1^q \equiv
\fr{q}{6\beta^2_q}\fr{V^{\prime 2}}{V^{1+q}},~~\delta_n \equiv
\frac{1}{H^n\dot{\phi}}\frac{d^{n+1}\phi}{dt^{n+1}}\approx
\delta_n^q,
\end{equation}
where they  satisfy the slow-roll condition:
$\epsilon_1<\xi,~|\delta_n|<\xi^n$ for a small parameter $\xi$.
 Here
 the subscript denotes  slow-roll (SR)-order in the slow-roll
calculation. A slow-roll parameter $\epsilon_1$ controls the
equation of states  $p=(-1+2\epsilon_1/3)\rho$ microscopically,
which implies that an accelerating expansion is always possible
for $\epsilon_1<1(p<-\rho/3)$\cite{Kinn}. The case of
$\epsilon_1=0$ corresponds to a de Sitter inflation. If one
chooses the inflation potential $V$ explicitly,  potential
slow-roll parameters ($\epsilon_1^q,~\delta_n^q$) will be found.

We  review how to calculate the cosmological parameters using the
slow-roll formalism. Introducing a new variable $u^q=a
(\delta\phi^q-\dot{\phi}\psi^q/H)$ where $\delta \phi^q (\psi^q)$
is a perturbed inflaton (perturbed metric), its Fourier modes
$u^q_k$ in the linear perturbation theory satisfies the Mukhanov
-type equation\footnote{We list the $q$-dependent potential:
$\frac{1}{z_q}\frac{d^2z_q}{d\tau^2}= 2a^2H^2\Big(
1+\epsilon_1+\frac{3}{2}\delta_1+ \frac{1}{q}\epsilon_1^2
+2\epsilon_1\delta_1 +\frac{1}{2}\delta_2\Big)$.  Here one change
occurs in coefficient of $\epsilon_1^2$: $1 \to 1/q$. Although the
 Gauss-Bonnet brane cosmology provides a complicated form,
its patch approximation provides  simple $q$-potentials with minor
change. This is the main reason why we choose  the patch cosmology
instead of the full Gauss-Bonnet brane cosmology.} \cite{muk}:
\begin{equation}
\label{eqsn} \frac{d^2u^q_k}{d\tau^2} +
\left(k^2-\frac{1}{z_q}\frac{d^2z_q}{d\tau^2}\right)u^q_k =
0.\end{equation} Here $\tau$ is the conformal time defined by
$d\tau=dt/a$ and $z_q=a\dot{\phi}/H$  is a parameter encoded all
information about a slow-roll inflation.

Before we proceed, we have to mention that Eq.(\ref{eqsn}) is the
nearly same form as in the conventional 4D perturbation
theory\cite{muk}. It  is well known that the perturbation theory
of braneworlds including Randall-Sundrum and Gauss-Bonnet models
is very different from the 4D-GR perturbation
theory\cite{MWBH,LT,RL,TL,HL,DLMS,LMW}. Making  use of the 4D
Mukhanov equation to study the braneworld perturbation, the
problem  is that this equation incorporates  4D metric (scalar)
perturbations only and thus there is no justification for using
this to describe the effect of  5D gravity on the brane. This
falls short of being a full 5D calculation as is required by the
braneworld sceranio. However, it was shown recently that even
though the effect of 5D metric perturbations on  inflation appears
to be large on small-scales (sub-horizon), on large-scales
(super-horizon) this effect is smaller than slow-roll corrections
to de Sitter background\cite{KLMW}. Further, the effect of 5D
metric perturbations is very small, at low energies, on
super-horizon and also this is suppressed, even at high energies,
on super-horizon. Therefore it is sensible to use the 4D
perturbation theory, especially for the study of GB and RS
 models in the high-energy regime because we always
compute the cosmological parameters on the super-horizon scale.

The  asymptotic solutions to Eq.(\ref{eqsn}) are obtained as
\begin{equation}\label{bc}
u^q_k \longrightarrow \left\{
\begin{array}{l l l}
\frac{1}{\sqrt{2k}}e^{-ik\tau} & \mbox{as} & -k\tau \rightarrow \infty \\
{\cal C}^q_k z_q & \mbox{as} & -k\tau \rightarrow 0.
\end{array} \right.
\end{equation}
 The first solution corresponds to a plane wave on
scale much smaller than the Hubble horizon of $d_H=1/H$
(sub-horizon), while the second is a growing mode on scale much
larger than the Hubble horizon (super-horizon). Using a relation
of  $R^q_{c{\bf k}}=-u^q_{\bf k}/z_q$ with $u^q_{\bf
k}(\tau)=a_{\bf k}u^q_k(\tau)+a^{\dagger}_{-{\bf
k}}u^{q*}_k(\tau)$ and a definition of
$P^q_{R_c}(k)\delta^{(3)}({\bf k}-{\bf
l})=\fr{k^3}{2\pi^2}<R^q_{c{\bf k}}(\tau)R^{q \dagger}_{c{\bf
l}}(\tau)>$, one finds the power spectrum for a curvature
perturbation on  the super-horizon scale
\begin{equation}\label{gps}
P^q_{R_c}(k) = \left(\frac{k^3}{2\pi^2}\right)
\lim_{-k\tau\rightarrow0}\left|\frac{u^q_k}{z_q}\right|^2 =
\frac{k^3}{2\pi^2}|{\cal C}^q_k|^2.
\end{equation}
Our task is to find ${\cal C}^q_k$ by solving the Mukhanov-type
equation (\ref{eqsn}). In general, it is not easy to find a
solution to this equation. However, we could solve it by using
either the slow-roll approximation \cite{SL} or the slow-roll
expansion\cite{SG}.
 In the slow-roll approximation  we take  $\epsilon_1$ and $\delta_1$ to be constant.
Thus this method  could not be considered as a general approach
beyond the first-order correction to the power
spectrum\cite{KLM,KLLM}. In order to calculate the power spectrum
up to second-order, one  uses the slow-roll expansion based on
Green's function technique. A key step is to account for a slowly
varying  nature of slow-roll parameters implied by two equations
$\dot{H}=-\frac{3}{2}q\beta^2_q \rho^{q-1}\dot{\phi}^2$ and
$\ddot{H}=2H\dot{H}[-(1-1/q)\epsilon_1 +\delta_1]$:
\begin{eqnarray} \label{slow-di}
\dot \epsilon_1
&=&2H\Big(\epsilon_1^2/q+\epsilon_1\delta_1\Big),~\dot{\delta}_1=
H(\epsilon_1\delta_1-\delta^2_1+\delta_2), \nonumber\\
\dot{\delta_2}&=&H(2\epsilon_1\delta_2-\delta_1\delta_2+\delta_3),~
\dot{\delta_3}=H(3\epsilon_1\delta_3-\delta_1\delta_3+\delta_4).
\end{eqnarray}
The above  means that the derivative of slow-roll parameters with
respect to time increases their SR-order by one.  After a lengthly
calculation following Ref.\cite{SG,KM2}, we obtain  the $q$-power
spectrum, $q$-spectral index, and $q$-running spectral index. See
Appendix A, for their explicit forms.

The tensor-to-scalar ratio $R_q$ is defined by \beq
R_q=16\fr{A_{T,q}^2}{A_{S,q}^2}.\eeq Here the $q$-scalar amplitude
in the leading-order is normalized by \beq
A_{S,q}^2=\fr{4}{25}P_{R_c}^{q,ESR} \eeq
 with the extreme slow-roll (ESR) power spectrum\footnote{Here, the extreme slow-roll
 limit means that $\epsilon_1 \to 0$ and $\delta_n \to 0$ in Eq.(\ref{2ndps}).
 Actually, this corresponds to the background spacetime of de Sitter inflation with $H$=const for the  slow-roll
 expansion.} \beq
 P_{R_c}^{q,ESR}=\fr{3q\beta^{2-\theta}_q}{(2\pi)^2}\fr{H^{2+\theta}}{2\epsilon_1}=
 \fr{1}{(2\pi)^2}\fr{H^4}{\dot{\phi}^2}.
 \eeq
The GR($q=1$) tensor amplitude up to leading-order is given by
\beq A_{T,GR}^2=\fr{1}{50}P_T^{ESR} \eeq where
$P_T^{ESR}=(2\kappa_4)^2\Big(\fr{H}{2\pi}\Big)^2$ since a tensor
can be expressed in terms of two scalars like $\delta \phi$. For
the GR case,  we could calculate  tensor spectra up to
higher-order as we wish to do  using the slow-roll expansion. On
the other hand, the tensor spectra for GB and RS models  are known
only for de Sitter brane like $a(t)\sim e^{Ht}$ \cite{LMW,DLMS}.
This means that tensor computation should be limited to the
leading-order in our calculation. These are given by \beq
A_{T,q}^2=A_{T,GR}^2F_{\beta}^2(H/\mu), \eeq where \beq
F_{\beta}^{-2}(x)=\sqrt{1+x^2}-\Big(\fr{1-\beta} {1+\beta}\Big)x^2
\sinh^{-1}\Big(\fr{1}{x}\Big).\eeq  In three different regimes, we
approximate $F_{\beta}^2$ as $F_q^2$: $F_1^2\approx
F_{\beta}^2(H/\mu \ll 1)=1$ for the  GR case ; $F_2^2 \approx
F_{\beta=0}^2(H/\mu \gg 1)=3H/(2\mu)$ for the RS case ; $F_{2/3}^2
\approx F_{\beta}^2(H/\mu \gg 1)=(1+\beta)/(2\beta)(\mu/ H)$ for
the GB case. The $q$-tensor amplitude up to leading-order is given
by \beq
 A_{T,q}^2=\fr{3q\beta^{2-\theta}_q}{(5\pi)^2}\fr{H^{2+\theta}}{2\zeta_q}
 \eeq
 with $\zeta_1=\zeta_{2/3}=1$ and $\zeta_2=2/3$\cite{Cal3}.
Finally, the tensor-to-scalar ratio is determined by \beq
\label{ttosr}
R_q=16\fr{A_{T,q}^2}{A_{S,q}^2}=16\fr{\epsilon_1}{\zeta_q}.\eeq
Considering a relation of $n_{T}^{q}=-(2+\theta)\epsilon_1$, one
finds that \beq
R_1=-8n_{T}^{1}=16\epsilon_1,~R_2=-8n_{T}^{2}=24\epsilon_1,~R_{2/3}=-16n_{T}^{2/3}=16\epsilon_1.\eeq
The above shows that the RS-consistency relation takes the same
form for GR case,  but the GB-consistency relation is different
from  those of RS and GR cosmological models.

 \section{Inflation with large-field potentials}

  First  we calculate  the
 scalar cosmological parameters
 for the large-field model using
 slow-roll expansion because there is  no upper-limit on this
 calculation.
 We choose  large-field potentials (LF) like $V(\phi)=V_0\phi^p$ with $p=2,4,6$ for our purpose.
 The LF-slow-roll parameters (V-SR), LF-power
spectrum, LF-spectral index, and LF-running spectral index are
given by Appendix B.   According to Ref.\cite{CT}, the quartic
potential is observationally disfavored when using the GR model
because the theoretical points are outside the 2$\sigma$ contour
bound for $45 \le N<60$. We note that   a range of $50 \le N \le
60$ is usually used  for a typical calculation of inflation. The
potentials of $V=V_0\phi^p$ with $p \ge 6$ are ruled out by the
CMB alone\cite{SDSS}. A potential of $V=V_0\phi^2$ is considered
as a promising one  because it is inside the 1$\sigma$ bound for
$50 \le N \le 60$. For the RS model, the quartic potential is
under strong observational pressure as is similar to the GR case
because it is outside the 2$\sigma$ bound for $45 \le N \le 60$.
The quadratic potential in the RS model is inside the 2$\sigma$
bound for $50 \le N \le 60$. The exponential potential is ruled
out for the RS case since it is far outside the 2$\sigma$ bound.
The quartic potential is ruled out for the GB model because it is
far outside the 2$\sigma$ bound, while the quadratic potential is
inside the 1$\sigma$ bound for $45 \le N \le 60$. It implies that
$V=V_0\phi^2$ is considered as a promising potential for GR and GB
models, while $V=V_0\phi^4$ is under significant pressure from
data. However, it does not mean that the quartic potential is
 ruled out by the current observation completely.

\begin{table}
\caption{The higher-order corrections to the spectral index
($n_s$) and running spectral index ($\alpha \equiv d n_s/d \ln
k$). Here we choose $N=50$ to find all theoretical parameters for
the large-field potentials (LF). For GB model with $p=6$, we have
to use the other result such as  the power-law inflation in Table
III.}
 \begin{tabular}{|c|c|c|c|c|}
   Patch & $p$ & Leading-order& First-order & Second-order\\\hline
         &     & $n_s$=0.9703 & 0.9700      & 0.9700  \\
         &2    & $\alpha = -0.0006$ & $-0.0006$ & $-0.0006$ \\ \cline{2-5}
     GB  & &     $n_s$= 0.9423 & 0.9393      & 0.9392 \\
$(q=2/3)$ &4   &  $\alpha = -0.0011$ & $-0.0013$ &$ -0.0013$\\
\cline{2-5}
              &     & N/A & N/A     & N/A   \\
         &6    & N/A & N/A  & N/A  \\ \hline

 &     & $n_s$=0.9604 & 0.9602      & 0.9602  \\
         &2    & $\alpha = -0.0008$ & $-0.0008$ & $-0.0008$ \\ \cline{2-5}
     GR  & &     $n_s$= 0.9412 & 0.9401      & 0.9401 \\
$(q=1)$ &4   &  $\alpha = -0.0012$ & $-0.0012$ &$ -0.0012$\\
\cline{2-5}
              &     & $n_s$=0.9223 & 0.9200      & 0.9200  \\
         &6    & $\alpha = -0.0015$ & $-0.0016$ & $-0.0016$  \\ \hline

                &     & $n_s$=0.9505 & 0.9503      & 0.9503  \\
         &2    & $\alpha = -0.001$ & $-0.001$ & $-0.001$ \\ \cline{2-5}
     RS  & &     $n_s$= 0.9408 & 0.9403      & 0.9404 \\
$(q=2)$ &4   &  $\alpha = -0.0012$ & $-0.0012$ &$ -0.0012$\\
\cline{2-5}
              &     & $n_s$=0.9360 & 0.9354      & 0.9354  \\
         &6    & $\alpha = -0.0013$ & $-0.0013$ & $-0.0013$
\end{tabular}
\end{table}

In this section we investigate this problem more carefully.  We
observe from Table II that the theoretical  parameters are not
significantly changed by the SR higher-order calculations. In
other words,  the theoretical parameters are insensitive to the
higher-order corrections.  It confirms that the leading-order
calculation is sufficient to compare with the observation data in
the slow-roll limit.  All of the running spectral indices are very
small when comparing with the WMAP data of
$dn_{s}/d\ln{k}=-0.031^{+0.016}_{-0.018}$ at $k_0=0.05 {\rm
Mpc}^{-1}$\cite{Wmap1}. Hence the large-field model is close to
the vanilla model with the zero-running spectral index\cite{SDSS}.

On the other hand, the theoretical parameters are significantly
changed by introducing different cosmological models. In testing
the large-field model with the data, a significant difference
appears between GB, GR and RS models. Here we mainly use the two
important parameters: the spectral index $n_s^q$ and $R_q$
obtained from the leading-order calculation: the first line in
Eq.(\ref{3rdnspi}) and Eq.(\ref{ttosr}) together with
$\epsilon_1^q$ in Eq.(\ref{pislp}). What we want to do is  to find
which $q$-model  moves theoretical points predicted by a given
potential quickly inside the 1$\sigma$ bound contour ($n_s^q \to
1$, $R_q \to 0$ and $d n_s/d \ln k \to 0$ inspired by the vanilla
model) for $50\le N \le 60$.

Consequently, the GB model is regarded as a promising one because
it accepts the quadratic potential clearly, but it rejects the
quartic potential because theoretical points are far outside the
$2\sigma$ bound. The GB model  improves the theoretical values
predicted by the GR case, whereas the RS model provides
indistinctive values more than the GR case. Actually the GB model
splits large-field potentials into three distinct regions clearly:
for $N=50$, $n_s^{GB}=0.97 \to 0.94~(p=2\to p=4),~R_{GB}=0.16 \to
0.61$ and the power-law inflation with $p=6$:
$n_s^{PI}=1-[(2r-1)/2r^4]^{1/3},~R_{PI}=16/r$ (see Table III).
Here, for the GB  model with $p=6$, we have to use the result of
the power-law inflation which corresponds to the border between
large-field and hybrid models. This implies that for the GB case,
$p=6$ no longer belong a class of large-field potentials. On the
other hand, we have $n_s^{GR}=0.96 \to 0.94 \to 0.92~(p=2\to
p=4\to p=6),~R_{GR}=0.16 \to 0.31 \to 0.47$, whereas
$n_s^{RS}=0.95 \to 0.94 \to 0.94,~R_{RS}=0.24 \to 0.32 \to 0.35$.
For $N=60$, $n_s^{GB}=0.98 \to 0.95~(p=2\to p=4),~R_{GB}=0.13 \to
0.52$ and the power-law inflation with $p=6$:
$n_s^{PI}=1-[(2r-1)/2r^4]^{1/3},~R_{PI}=16/r$. Also we obtain
$n_s^{GR}=0.97 \to 0.95 \to 0.93~(p=2\to p=4\to p=6),~R_{GR}=0.13
\to 0.26 \to 0.39$, while  $n_s^{RS}=0.96 \to 0.95 \to
0.95,~R_{RS}=0.20 \to 0.26 \to 0.30$. The theoretical points
predicted by the RS model lie very close to the border between the
regions allowed and disallowed by the observation.

As is shown in Table III, three different potentials  give the
 nearly same power-law inflation when the patch cosmology is used to
 calculate the cosmological paramters\cite{KM2}.
We find the allowed steps from the observationally favored
potential to the power-law inflation which corresponds to the
border between large-field and hybrid models: for the GB case,
$\phi^2 \to \phi^4 \to \phi^6$ (three steps), for the GR case,
$\phi^2 \to \phi^4 \to \phi^6 \to e^{-\phi}(\approx \phi^p,p\to
\infty)$ (four steps), and for the RS case, $\phi^2 \to \phi^4 \to
\phi^6 \to e^{-\phi} \to \phi^{-2}$ (five steps).

\begin{table}
 \caption{Potentials, spectral index,  running spectral index, and tensor-to-scalar ratio for a power-law
 inflation with $a(t)=a_0t^r$ with $r>1$. Three different
 potentials give the nearly same result when
  the patch cosmology is applied to calculate cosmological parameters.}

\begin{tabular}{c|c|c|c|c}
  model & $V^{PI}$ & $dn_s^{PI}$
   &$d n^{PI}_s/d\ln k$ & $ R^{PI}$\\
  \hline
  GB & $V_0\phi^6 $
&
$1-\Big[\fr{2r-1}{2r^4}\Big]^{\fr{1}{3}}-\Big[\fr{2r-1}{2r^4}\Big]^{\fr{2}{3}}-\Big[\fr{2r-1}{2r^4}\Big]$
& 0 & $16/r$ \\
GR &$V_0 \exp(-\sqrt{2\kappa^2_4/r}~\phi)$
&$1-\fr{2}{r}-\fr{2}{r^2}-\fr{2}{r^3}$ &0 & $16/r$\\
  RS &$V_0\phi^{-2}$
&$1-3\Big[\fr{6}{6r-1}\Big]-3\Big[\fr{6}{6r-1}\Big]^{2}-3\Big[\fr{6}{6r-1}\Big]^{3}$&
0 & $24/r$

\end{tabular}

 \end{table}

\section{discussions}
First of all, we mention that Eq.(\ref{eqsn}) incorporates 4D
metric (scalar) perturbations in the cosmological perturbation
theory and thus there is no justification for using this to
describe the effect of 5D gravity on inflation confined to the
brane. However, it was shown  recently that at low energies, the
effect of 5D metric perturbations is very small on super-horizon
and  even at high energies, this is suppressed on super-horizon.
Hence  it is not serious to use the 4D cosmological  perturbation
theory, especially for the study of GB and RS cosmological models
in the high-energy regime because we always compute  the
cosmological parameters on the super-horizon scale.

Since second-order corrections are rather small in the slow-roll
limit, their theoretical points are not significantly moved from
the leading-order results. Hence we use the results from the
leading-order calculation to compare the theoretical parameters
with the data.  It turns out that the GB cosmological model is a
 promising model  to discriminate between the quadratic potential and
quartic potential, when comparing with the observation data. This
is mainly because the GB model  divides  large-field potentials
into three distinct regions in the $n_s^{q}$-$R_q$ plane clearly:
$p=2$ (inside the $1\sigma$ bound), $p=4$ (far outside the
$2\sigma$ bound), $p=6$ ( the border between large-fields and
hybrid models). On the other hand, the GR (RS) models  require
four (five) regions in the $n_s^{q}$-$R_q$ plane. The GB model
provides the tightest constraint on the large-field model, whereas
the RS model provides the loosest constraint because the
theoretical points are too close to distinguish between  quadratic
potential and quartic potential. In this sense, the
Randall-Sundrum braneworld in the high-energy regime (the RS
model) is regarded to be the worst case.

Consequently, we show that it is easier to discriminate between
quadratic and quartic inflation potentials in the GB cosmological
model rather than the GR or RS cosmological models.

\subsection*{Acknowledgements}
We thank  Hungsoo Kim, H. W. Lee and G. Calcagni for helpful
discussions. Y.S. was supported in part by KOSEF, Project No.
R02-2002-000-00028-0. K.H. was in part supported by KOSEF,
Astrophysical Research Center for the Structure and Evolution of
the Cosmos.

\subsection* {Appendix A: Second-order corrections to
cosmological parameters}

The  $q$-power spectrum is calculated as
\begin{eqnarray}
 \label{2ndps}
P^{q}_{R_c}(k) & = &  \fr{H^4}{(2\pi)^2\dot{\phi}^2} \left\{ 1
-2\epsilon_1 + 2\alpha(2\epsilon_1+\delta_1) \right. \\
&& \left.
+\left((8-4/q)\alpha^2-4(1-1/q)\alpha-(19+4/q)+(2+1/3q)\pi^2\right)\epsilon_1^2 \right. \nonumber \\
&& \left. +
\left(3\alpha^2+2\alpha-22+29\pi^2/12\right)\epsilon_1\delta_1 +
\left(3\alpha^2-4+5\pi^2/12\right)\delta_1^2 -
\left(\alpha^2-\pi^2/12\right)\delta_2 \right\} \nonumber
\end{eqnarray}
and the right hand side should be evaluated at horizon crossing of
$k=aH$. $\alpha$ is defined by $\alpha=2-\ln2-\gamma \simeq
0.7296$, where $\gamma$ is the Euler-Mascheroni constant with
$\gamma \simeq 0.5772.$  We note that the $q$-dependent terms
appear in the second-order corrections. Using a relation of $d\ln
k \simeq Hdt$ and Eq.(\ref{slow-di}), the $q$-spectral index
defined by
\begin{equation}
n_{s}^q(k) = 1 + \frac{d \ln P^{q}_{R_c}}{d \ln k}
\end{equation}
can be calculated as

\begin{eqnarray}
\label{11} n^{q}_{s}(k) =& & 1 - 4\epsilon_1 - 2\delta_1
+(-4-4/q+8\alpha/q)\epsilon_1^2
              + (10\alpha -6)\epsilon_1\delta_1 -2\alpha\delta_1^2+2\alpha\delta_2
              \\ \nonumber
          & & + \Big\{
                    -16\alpha^2/q^2+(16/q^2+24/q)\alpha-4-16/q^2-88/q +
                    (4/3q^2+8/q)\pi^2
                \Big\}\epsilon_1^3       \\ \nonumber
          & &     +\Big\{
                    -(26/q+5)\alpha^2+(32+28/q)\alpha-112-60/q+(125/12+37/6q)\pi^2
                \Big\}\epsilon_1^2\delta_1
                \\ \nonumber
          & & + \left(
                    -3\alpha^2+4\alpha-30+13\pi^2/ 4
                \right)\epsilon_1\delta_1^2
                +\left(
                    -7\alpha^2+8\alpha-22+31\pi^2/12
                \right)\epsilon_1\delta_2
                \\ \nonumber
          & & + \left(
                    -2\alpha^2+8-5\pi^2/6
                \right)\delta_1^3
                +\left(
                    3\alpha^2-8+3\pi^2/4
                \right)\delta_1\delta_2
                +\left(
                    -\alpha^2+\pi^2/12
                \right)\delta_3.
\end{eqnarray}
The $q$-running spectral index  is determined by

\begin{eqnarray}
\label{12} \frac{d}{d\ln k} n^{q}_{s} & = &
-8\epsilon^{2}_{1}/q-10\epsilon_1\delta_1+2\delta^{2}_{1}-2\delta_2
\\ \nonumber
&+& \left(-16/q^2-24/q+32\alpha/q^2\right)\epsilon^{3}_{1}
\\ \nonumber
&+& \left(-32 -28/q+(10+52/q)\alpha\right)\epsilon^{2}_{1}\delta_1
\\ \nonumber
&+&(6\alpha-4)\epsilon_1\delta^{2}_{1}+(14\alpha
-8)\epsilon_1\delta_2+4\alpha\delta^{3}_{1}-6\alpha\delta_1\delta_2+2\alpha\delta_3 \\
\nonumber
         &+&\Big\{-96\alpha^2/q^2+(96/q^3+176/q^2)\alpha-(96/q^3+544/q^2+48/q) +
         (8/q^3+48/q^2)\pi^2\Big\}\epsilon_1^4
         \\ \nonumber
         &+& \Big\{-(200/q^2+46/q+5)\alpha^2+(208/q^2+352/q+42)\alpha \Big\}\epsilon_1^3\delta_1 \\ \nonumber
         &+&\Big\{-(336/q^2+1064/q +168)+(98/3q^2+575/6q+125/12)\pi^2\Big\}\epsilon_1^3\delta_1
               \\ \nonumber
         &+&\Big\{-(84/q+21)\alpha^2+(92/q+100)\alpha-(240/q+400)+(25/q+151/4)\pi^2
             \Big\}\epsilon_1^2\delta_1^2  \\ \nonumber
         &+&\Big\{-(40/q+19)\alpha^2+(44/q+62)\alpha-(104/q+164)+(34/3q+187/12)\pi^2
         \Big\}\epsilon_1^2\delta_2
               \\ \nonumber
         &+&\left(-6\alpha^2+4\alpha+24-5\pi^2/
             2\right)\epsilon_1\delta_1^3
            +\left(-4\alpha^2+10\alpha-106+34\pi^2/
             3\right)\epsilon_1\delta_1\delta_2
               \\ \nonumber
         &+&\left(-10\alpha^2+10\alpha-22+17\pi^2/
             6\right)\epsilon_1\delta_3
            +\left(6\alpha^2-24+5\pi^2/2\right)\delta_1^4
               \\ \nonumber
         &+&\left(-12\alpha^2+40-4\pi^2\right)\delta_1^2\delta_2
            +\left(4\alpha^2-8+2\pi^2/3\right)\delta_1\delta_3
               \\ \nonumber
         &+&\left(3\alpha^2-8+3\pi^2/4\right)\delta_2^2
            +\left(-\alpha^2+\pi^2/12\right)\delta_4.
            \nonumber
\end{eqnarray}

\subsection*{Appendix B: Cosmological parameters for large-field
potentials}

The LF slow-roll parameters are determined by

\begin{eqnarray}\label{pislp}
\epsilon_1^q &=& {{qp} \over 2}{1 \over X}, \nonumber \\
\delta_1^q &=& {1 \over 2}{(2 - 2p + qp)  \over X}, \nonumber \\
\delta_2^q &=& {1 \over 2}{(2 - 2p + qp)(3 - 2p + qp)
 \over X^2}, \nonumber \\
\delta_3^q &=& {1 \over 4}{(2 - 2p + qp)(3 - 2p + qp)(10 - 6p +3qp) \over X^3 }, \nonumber \\
\delta_4^q &=& {1 \over 4}{(2 - 2p + qp)(3 - 2p + qp)(10 - 6p
+3qp)(7 - 4p + 2qp) \over X^4}.
\end{eqnarray}
with $X \equiv [(q-1)p + 2]N + {{qp} \over 2}$. Substituting these
into Eqs.(\ref{2ndps}), (\ref{11}) and (\ref{12}), one finds the
corresponding inflation parameters.
 The
LF-power spectrum takes the form
\begin{eqnarray}\label{2ndpssb}
P_{Rc}^{LF} &=& {\left({{H^2}\over{2\pi\phi}} \right)}^2 \Big\{ 1
+ \Big[(2 - 2p + 3qp)\alpha - qp\Big]{1 \over X} + \Big[((3q^2 -
7q/2 + 1)p^2
 + (2q - 1)p)\alpha^2   \nonumber \\
           & +& \alpha(-q^2 p^2/ 2 + qp) +5 \pi^2 q^2p^2/ 4 -45 q^2p^2/4 - 41
\pi^2
qp/24 + 11 \pi^2 qp/6  \nonumber \\
           &+& 14qp^2 - 15qp+ 7 \pi^2p^2/12 -4p^2 -5 \pi^2p/4 + 8p + 2\pi^2/3 -
4 \Big]{1 \over X^2} \Big\}.
\end{eqnarray}

The LF-spectral index is given by
\begin{eqnarray}\label{3rdnspi}
n(k)_{s}^{LF} &=&1- \left\{\begin{array}{cl}
                  (3q-2)p+2
                   \end{array}\right\}
                   \frac{1}{X}\\ \nonumber
                   &+&\left\{\begin{array}{cl}
                    \alpha((3q^2-5q+2)p^2+(8q-6)p+4))-
                    (5q^2/2-2q)p^2-3qp
                   \end{array}\right\}
                   \frac{1}{X^{2}}\\ \nonumber
                   &+&\left\{\begin{array}{cl}
                    (-3q^3+8q^2-7q+2)p^3-(14q^2-24q+7q+10)p^2-(20q-16)p-8
                   \end{array}\right\}
                   \frac{\alpha^2}{X^{3}}\\ \nonumber
                   &+&\left\{\begin{array}{cl}
                    (13q^3/2-23q^2/2+5q)p^3+(20q^2-17q)p^2
                    +14qp
                   \end{array}\right\}
                   \frac{\alpha}{X^{3}}\\ \nonumber
                   &+&\left\{\begin{array}{cl}
                  (5\pi^2/2-99/4)q^3-(71\pi^2/12+105/2)q^2
                 +(55\pi^2/12-36)q -7\pi^2/6+8
                   \end{array}\right\}
                   \frac{p^3}{X^{3}}\\ \nonumber
                   &+&\left\{\begin{array}{cl}
                  (26\pi^2/3-157/2)q^2
                 +(-13\pi^2+102)q+ 29\pi^2/6+32
                   \end{array}\right\}
                   \frac{p^2}{X^{3}}\\ \nonumber
                   &+&\left\{\begin{array}{cl}
                   (-102+13\pi^2)q-32+29\pi^2/6)
                   \end{array}\right\}
                   \frac{p^2}{X^{3}}\\ \nonumber
                   &+&\left\{\begin{array}{cl}
                   [(-68+26\pi^2/3)q+(40-19\pi^2/3)]p-16+8\pi^2/3  \\
                   \end{array}\right\}
                   \frac{1}{X^{3}}.
\end{eqnarray}
The first line corresponds to the leading-order calculation.
Finally, the LF-running spectral index is found to be
\begin{eqnarray} \label{rsi4th}
 \frac{d n^{LF,q}_s}{d\ln k}&=&-\left\{\begin{array}{cl}
                                                  (3q^{2}-5q+2)p^{2}
                                                  +(8q-6)p+4
                                                \end{array}\right\}
                                                \frac{1}
                                 {X^2}\\ \nonumber
                            &+& \left\{\begin{array}{cl}
                                      (6q^3-16q^2+14q-4)p^3+(28q^2-48q+20)p^2
                                      +(40q-32)p+16
                                          \end{array}\right\}
                                    \frac{\alpha}{X^3}\\\nonumber
                                    &+& \left\{\begin{array}{cl}
                                          (-13q^{3}/2+23q^{2}/2-5q)p^{3}
                                          -(20q^{2}-17q)p^{2}-14qp
                                          \end{array}\right\}
                                    \frac{1}{X^3}\\\nonumber
                             &+&\left\{\begin{array}{cl}
                               (-9q^{4}+33q^{3}-45q^{2}+27q-6)p^{4}
                               -(60q^{3}-162q^{2}+144q-42)p^{3}\\
                               -(144q^{2}-252q+108)p^{2}
                               -(144q-120)p-48
                             \end{array}\right\}\frac{\alpha^{2}}
                                   {X^4}\\\nonumber
                              &+& \left\{\begin{array}{cl}
                              (45q^{4}/2-62q^{3}+133q^{2}/2-17q)p^{4}
                                         +(113q^{3}-204q^{2}+91q)p^{3}\\
                                         +(182q^{2}-160q)p^{2}+92qp
                              \end{array}\right\}
                              \frac{\alpha}{X^4}\\\nonumber
                              &+& \left\{\begin{array}{cl}
                              (-155/2+15\pi^{2}/2)q^{4}
                              +(475/2-101\pi^{2}/4)q^{3}
                              -(268-63\pi^{2}/2)q^{2}\\
                              +(132-69\pi^{2}/4)q-24+7\pi^{2}/2
                              \end{array}\right\}
                              \frac{p^{4}}{X^4}\\\nonumber
                              &+& \left\{\begin{array}{cl}
                              (-394+41\pi^{2})q^{3}
                              +(865-201\pi^{2}/2)q^{2}
                              -(618+81\pi^{2})q+144\\
                              -43\pi^{2}/2
                              \end{array}\right\}
                              \frac{p^{3}}{X^4}\\\nonumber
                              &+ &\left\{\begin{array}{cl}
                              (-682+78\pi^{2})q^{2}
                              +(936-123\pi^{2})q
                              -(312-48\pi^{2})
                              \end{array}\right\}
                              \frac{p^{2}}{X^4}\\\nonumber
                              &+& \left\{\begin{array}{cl}
                              (-456+60\pi^{2})q-(288+46\pi^{2})\\
                              \end{array}\right\}
                              \frac{p}{X^4}\\\nonumber
                              &+& \left\{\begin{array}{cl}
                             -96+16\pi^{2}\\
                              \end{array}\right\}
                              \frac{1}{X^4}.
\end{eqnarray}
The first line is  the leading-order calculation.

\end{document}